\documentclass[journal,10pt,twocolumn]{IEEEtran}
\usepackage{amsfonts}
\IEEEoverridecommandlockouts

\ifCLASSINFOpdf
\else
\fi
\usepackage{float}
\usepackage{epsfig}
\usepackage{graphicx}
\usepackage{subfigure}
\usepackage{psfig}
\usepackage{epsf}
\usepackage[cmex10]{amsmath}
\usepackage{booktabs}
\usepackage{fancyhdr}
\usepackage{slashbox}
\usepackage{caption2}
\usepackage{color}

\begin{document}
\title{FFDNet-Based Channel Estimation for Massive MIMO Visible Light Communication Systems}
\author{{\IEEEauthorblockN{Zhipeng Gao, Yuhao Wang}, \emph{Senior Member, IEEE}, Xiaodong Liu, \emph{Student Member, IEEE}, \\ Fuhui Zhou, \emph{Member, IEEE}, Kai-Kit Wong, \emph{Fellow, IEEE}}
\thanks{Manuscript received Jul. 25, 2019; revised Oct 29, 2019 and accepted Nov. 16, 2019. Date of publication ****; date of current version ****. This work was supported in part by the National Natural Science Foundation of China (61661028 and 61701214), in part by the Excellent Youth Foundation of Jiangxi Province under Grant 2018ACB21012. The associate editor coordinating the review of this paper and approving it for publication was Yue Gao. (Corresponding author:  Yuhao Wang and Fuhui Zhou.)

Z. Gao and Y. Wang are with Nanchang University, China (e-mail: gdpeungee@email.ncu.edu.cn, wangyuhao@ncu.edu.cn).

F. Zhou is with the College of Electronic and Information Engineering,
Nanjing University of Aeronautics and Astronautics, Nanjing, 210000, P. R.
China. (e-mail: zhoufuhui@ieee.org).

X. Liu is with Wuhan University, China (e-mail: xiaodongliu@whu.edu.cn).

K. K. Wong is with University College London, UK (e-mail: kai-kit.wong@ucl.ac.uk).}}
\maketitle
\begin{abstract}
Channel estimation is of crucial importance in massive multiple-input multiple-output (m-MIMO) visible light communication (VLC) systems. In order to tackle this problem, a fast and flexible denoising convolutional neural network (FFDNet)-based channel estimation scheme for m-MIMO VLC systems was proposed. The channel matrix of the m-MIMO VLC channel is identified as a two-dimensional natural image since the channel has the characteristic of sparsity. A deep learning-enabled image denoising network FFDNet is exploited to learn from a large number of training data and to estimate the m-MIMO VLC channel. Simulation results demonstrate that our proposed channel estimation based on the FFDNet significantly outperforms the benchmark scheme based on minimum mean square error.
\end{abstract}
\begin{IEEEkeywords}
Channel estimation, m-MIMO, visible light communication, FFDNet, deep learning.
\end{IEEEkeywords}
\IEEEpeerreviewmaketitle
\section{Introduction}
\IEEEPARstart{W}{ITH} the ever increasing mobile data traffic, massive multiple-input multiple-output (m-MIMO) is promising to enhance the communication capacity and the spectrum efficiency of the existing visible light communication (VLC) systems \cite{R. Wang}-\cite{J. Zhang}. However, such benefits may be limited due to the inaccuracy of channel estimation \cite{F. Rusek}. Thus, an accurate channel estimation is of crucial importance in the m-MIMO VLC systems. However, when the dimensions of the channel matrix become large, it is extremely challenging to design efficient channel estimation techniques and achieve an accurate channel estimation for the m-MIMO systems \cite{J. Yang}-\cite{C. Huang}. Moreover, the conventional channel estimation techniques for m-MIMO systems such as minimum mean square error (MMSE) are typically dependent on the statistical properties of the channel and specific prior information \cite{M. Morelli}-\cite{X. Chen}. In practice, it is difficult to obtain those statistical properties, especially when the number of transmitters and that of receivers are large. Therefore, the performance of those conventional channel estimation methods is not always optimal.

In order to solve these problems, deep learning (DL) has attracted an increasing attention and has great potential to be applied for channel estimation of the wireless communication systems \cite{J. Gao}-\cite{H. He}. Specifically, in \cite{J. Gao}, a DL-based signal detector without requiring prior knowledge about the channel state information or background noise was proposed for cooperative detection. And in \cite{H. Huang}, a DL-based scheme was proposed to achieve the super-resolution channel estimation and direction-of-arrival estimation in m-MIMO systems. Since the line-of-sight (LOS) link communication is dominant in VLC systems, the m-MIMO VLC channel has the characteristic of sparsity. This is similar to that in \cite{H. He} where the mmWave m-MIMO channel matrix was regarded as a two-dimensional (2D) noise-free natural image by exploiting the sparsity of the channel. Then a convolutional neural network (CNN)-based image denoising network was exploited to estimate the m-MIMO mmWave channel. Motivated by those facts, the m-MIMO VLC channel matrix can be also identified as a 2D noise-free natural image.

Based on the above-mentioned analyses, it is feasible to utilize the theory of image denoising in DL for estimating the m-MIMO VLC channel. However, the state-of-the-art CNN-based image denoising methods are still limited in flexibility and efficiency \cite{K. Zhang}. To overcome the drawbacks of the existing CNN-based image denoising methods, a image denoising network called fast and flexible denoising convolutional neural network (FFDNet) was proposed in \cite{K. Zhang}. FFDNet has the superiority in terms of both denoising performance and computation efficiency. Moreover, FFDNet performs effectively and flexibly on the denoising of images that are corrupted by additive white gaussian noise (AWGN). Therefore, it is promising to apply FFDNet into the m-MIMO VLC systems for realizing the channel estimation.

Motivated by the above-mentioned facts, a channel estimation scheme is proposed based on FFDNet in a m-MIMO VLC system. The main contributions of our work are summarized as follows. A m-MIMO VLC system using a transceiver array which is composed of massive LEDs and PDs is considered. Then, the concept of image denoising is exploited and a trained FFDNet network is used to estimate the m-MIMO VLC channel. Moreover, the performance of channel estimation using FFDNet is evaluated and compared with that obtained by the traditional MMSE channel estimation scheme. Simulation results demonstrate that the performance of FFDNet with a fixed input noise level is better than that of MMSE at high real noise levels. Furthermore, the performance gain can be further improved by FFDNet with a tunable input noise level.

The rest of this paper is organized as follows. Section II comprehensively presents the m-MIMO VLC systems and the architecture of FFDNet image denoising network. In Section III, simulation results are presented to evaluate the scheme. Finally, the paper concludes with Section IV.

\section{Channel and FFDNet Network}
\subsection{System Scenario and Channel Model}
\begin{figure}[!htbp]
\centering
\includegraphics[height=1.7 in,width=2.0 in]{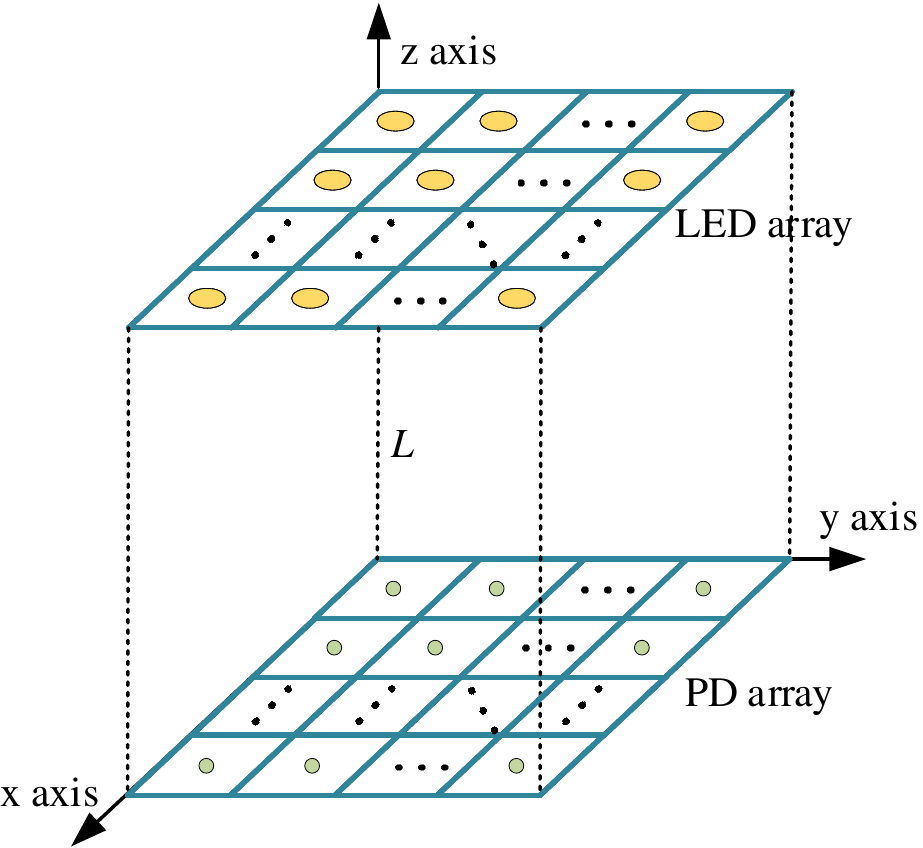}
\caption{The m-MIMO VLC system model.} \label{fig.1}
\end{figure}
In this paper, a m-MIMO VLC system is considered with ${N_t}$ light emitting diodes (LEDs) as the transmitters and ${N_r}$ photo detectors (PDs) as the receivers. It is assumed that the specifications and performance of all LEDs are identical, as are the PDs. As shown in Fig. 1, the LED-array plane is parallel to the PD-array plane, and all LEDs and PDs are arranged with an equal interval \cite{K. Xu}. And the vertical distance between the LED-array and the PD-array is ${L}$.
\begin{figure}[!htbp]
\centering
\includegraphics[height=1.6 in,width=1.7 in]{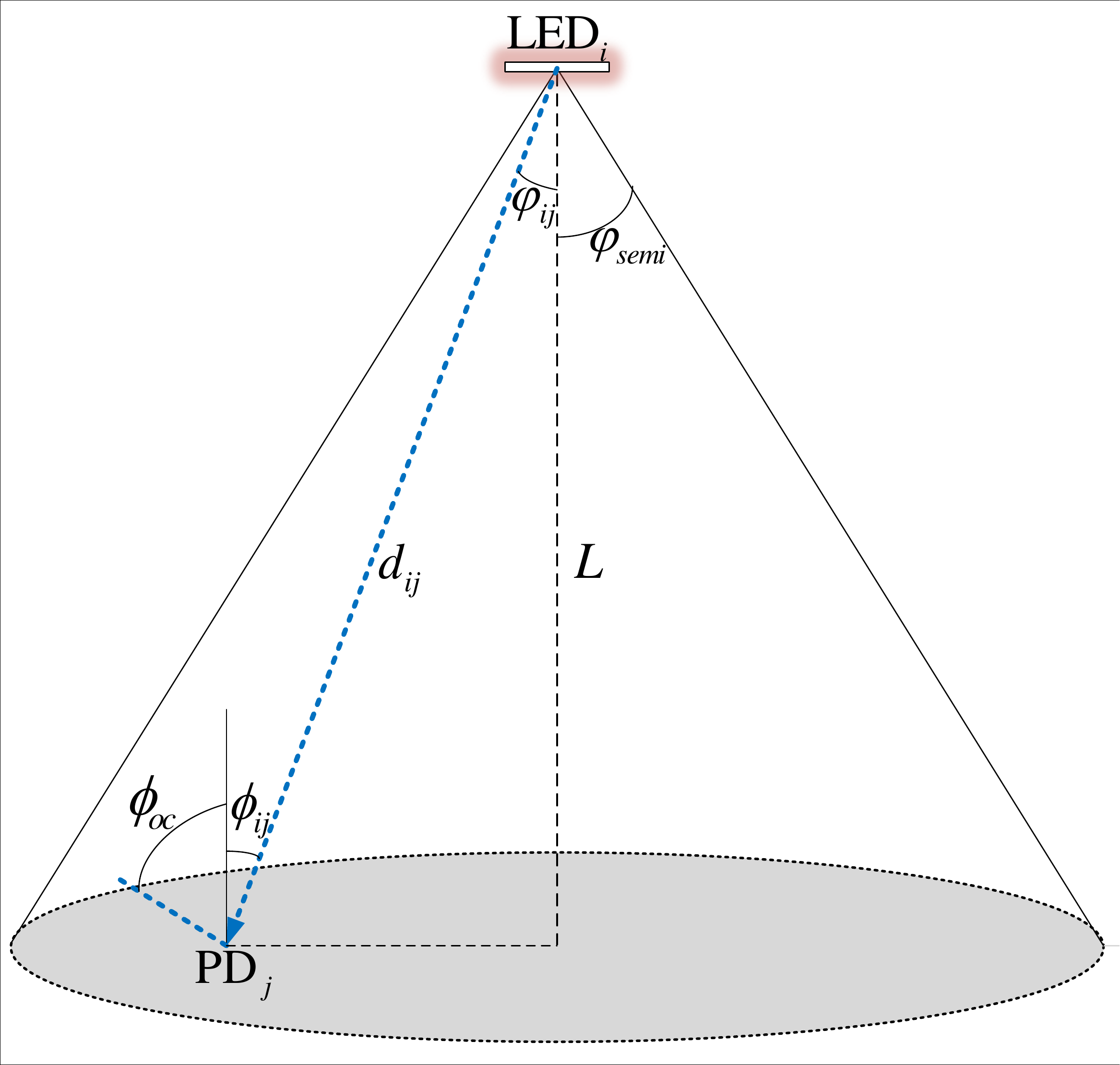}
\caption{The VLC channel model.} \label{fig.2}
\end{figure}

The current settings of the m-MIMO VLC channel is modeled as the LOS as shown in Fig. 2, since multipath delays resulting from reflections and defuse refractions are typically negligible in m-MIMO VLC scenarios \cite{P. C. Sofotasios}. The channel gain ${h_{ij}}$ between the $i{\rm{th}}$ LED and the $j{\rm{th}}$ PD is given by \cite{L. Yin}
\begin{align}
{h_{ij}} = \frac{{{A_r}}}{{{d_{ij}}^2}}{R_o}\left( {{\varphi _{ij}}} \right){T_s}\left( {{\phi _{ij}}} \right)g\left( {{\phi _{ij}}} \right)\cos \left( {{\phi _{ij}}} \right),
\end{align}
where $i = 1,2,3,...,N_t$ and $j = 1,2,3,...,N_r$. ${{A_r}}$ represents the effective collection area of the receiver PD. ${{d_{ij}}}$ denotes the distance between the $i{\rm{th}}$ transmitting LED and the $j{\rm{th}}$ receiving PD. ${{\phi _{ij}}}$ and ${{\varphi _{ij}}}$ represent the incidence angle and the irradiance angle, respectively. ${T_s}\left( {{\phi _{ij}}} \right)$ is the gain of the optical filter, and $g\left( {{\phi _{ij}}} \right)$ denotes the gain of the optical concentrator (OC), which can be calculated by
\begin{align}
g\left( {{\phi _{ij}}} \right) = \left\{ {\begin{array}{*{20}{c}}
{\frac{{{n^2}}}{{{{\sin }^2}\left( {{\phi _{oc}}} \right)}}},&{{\rm{if \quad }}0 \le {{\phi _{ij}}} \le {\phi _{oc}}},\\
0,&{{\rm{otherwise \quad }}{{\phi _{ij}}} > {\phi _{oc}}},
\end{array}} \right.
\end{align}
where ${{\phi _{oc}}}$ is the field of view (FOV) of the PD. $n$ denotes the refractive index of OC. Moreover, ${R_o}\left( {{\varphi _{ij}}} \right)$ represents the Lambertian radiant intensity of the transmitting LEDs, which can be expressed as
\begin{align}
{R_o}\left( {{\varphi _{ij}}} \right) = \frac{{m + 1}}{{2\pi }}{\cos ^{m}}\left( {{{\varphi _{ij}}}} \right),
\end{align}
where $m$ is the order of Lambertian emission, which is given by
\begin{align}
m =  - \frac{{\ln \left( 2 \right)}}{{\ln \left( {\cos \left( {{\varphi _{semi}}} \right)} \right)}},
\end{align}
where ${{\varphi _{semi}}}$ represents the LED transmitter semi-angle at a half power level.

Motivated by the work in \cite{H. He}, after obtaining the channel matrix ${\bf{H}}$ of the m-MIMO VLC system, the channel matrix can be treated as a noise-free channel image $x$ with a size $w \times l$. And it becomes a noisy channel image $y$ of size $w \times l$ after it is corrupted by AWGN with real noise level ${\sigma _{\rm{o}}}$. Then the noisy channel image $y$ is input to a trained FFDNet network to estimate the channel.
\subsection{The FFDNet Architecture}
\begin{figure*}[!htbp]
\centering
\includegraphics[height=1.1 in,width=6.8 in]{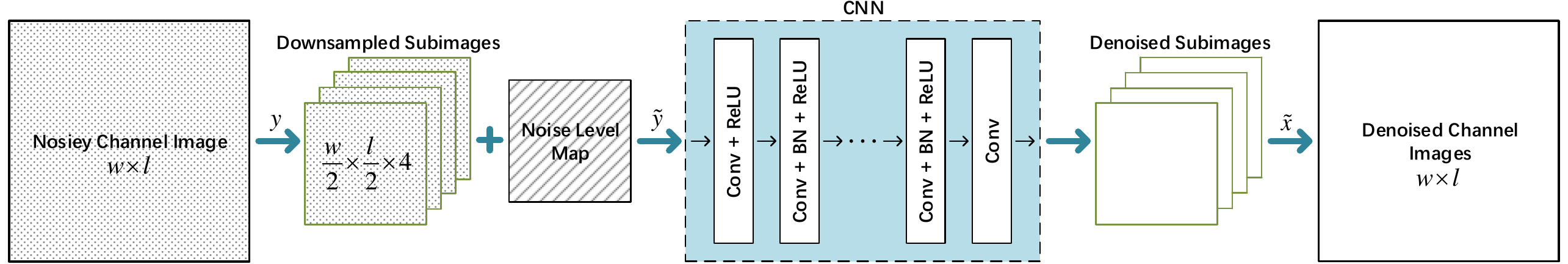}
\caption{The system architecture of FFDNet.} \label{fig.3}
\end{figure*}
As shown in Fig. 3 at the top of the next page, the first layer of FFDNet is a reversible downsampling process which reshapes the noisy channel image $y$ into four downsampled sub-images with size $\frac{w}{2} \times \frac{l}{2} \times 4$. The downsampling process can significantly improve the training speed without reducing the modeling ability. Moreover, different from the denoising convolutional neural network (DnCNN), the denoising on downsampled sub-images can effectively increase the receptive field without employing the dilated convolution and lead to a moderate network depth \cite{K. Zhang}. After the operation of downsampling, a tunable noise level map $\bf{M}$ with the input noise level\footnote{$\sigma $ is the noise level of the noise level map $\bf{M}$ which is used to control the trade-off between noise reduction and detail preservation. ${\sigma _{\rm{o}}}$ is the noise level of the AWGN that added to the channel matrix of the m-MIMO VLC system.} $\sigma $ is combined with the downsampled sub-images to establish $\widetilde y$ with a size $\frac{w}{2} \times \frac{l}{2} \times \left( {4 + 1} \right)$ as the input of the CNN.

With the tensor $\widetilde y$ as input, the following layer is a CNN with the depth ${D_e}$. As shown in Fig. 3, in order to effectively speed up the training process and improve the denoising performance of FFDNet, each layer in the CNN is composed of a specific combination of three types of operations: convolution (Conv), rectified linear units (ReLU) and batch normalization (BN) \cite{K. Zhang}. The first convolution layer ``Conv+ReLU" uses 64 filters with a size $3 \times 3$ to generate 64 feature maps, and ReLU is adopted for nonlinearity. The middle layers ``Conv+BN+ReLU" adopt 64 filters with a size $3 \times 3 \times 64$ and BN is added between the operations of convolution and the ReLU. And the last layer ``Conv" uses a filter with the size $3 \times 3 \times 64$ to reconstruct the output. Furthermore, after each convolution, zero-padding is employed to guarantee that the size of the feature maps is not changed. After the CNN, an upscaling operation is exploited as the reverse process of the downsampling process applied in the input stage to produce the denoised channel image $\widetilde x$ of size $w \times l$.
\subsection{The Noise Level Map}
Most of the model-based denoising methods aim to solve the problem, given as \cite{K. Zhang}
\begin{align}
\widetilde x = \arg {\min _x}\frac{1}{{2{\sigma ^2}}}{\left\| {y - x} \right\|^2} + \lambda \Phi \left( x \right),
\end{align}
where $\frac{1}{{2{\sigma ^2}}}{\left\| {y - x} \right\|^2}$ is the data fidelity term with the input noise level $\sigma $, and
$\Phi \left( x \right)$ represents the regularization term associated with image prior. $\lambda $ is used to control the balance between the data fidelity term and the regularization term. In order to solve this problem, an implicit function can be defined as \cite{K. Zhang}
\begin{align}
\widetilde x = F\left( {y,\sigma,\lambda;\Theta } \right),
\end{align}
where $\Theta $ is the trainable parameters in CNN. Since $\lambda $ can be equal to $\sigma $, eq. (6) can be rewritten as
\begin{align}
\widetilde x = F\left( {y,\sigma;\Theta } \right).
\end{align}

In this case, the input noise level $\sigma $ can also control the balance between the data fidelity term and the regularization term. And these model-based methods are flexible in image denoising by setting various input noise levels. Thus it is promising to exploit CNN to learn the explicit mapping of eq. (7) which takes $y$ and $\sigma $ as inputs. However, $y$ and $\sigma $ cannot be fed into CNN directly since the dimensions of them are different. Therefore, in the FFDNet, a tunable noise level map $\bf{M}$ which has the same dimension with $y$ is utilized. And all the elements of $\bf{M}$ are $\sigma $. Then the implicit function can be further written as
\begin{align}
\widetilde x = F\left( {y,\bf{M};\Theta } \right).
\end{align}
\subsection{Dataset Generation and Network Training}
In order to train the FFDNet network, it is necessary to obtain ${N_p}$ training data $\left\{ {\left( {{y_k},\bf{M_k};\bf{H_k}} \right)} \right\}_{k = 1}^{N_p}$. Firstly, we obtain the channel matrix $\bf{{H_k}}$ by implementing the m-MIMO VLC channel and treat it as a noise-free channel image. Secondly, the noise-free channel image $\bf{{H_k}}$ is added with AWGN that has zero mean and variance ${\left( {{{{\sigma _{\rm{o}}}} \mathord{\left/
 {\vphantom {{{\sigma _k}} {255}}} \right.
 \kern-\nulldelimiterspace} {255}}} \right)^2}$ to form the noisy channel image ${{y_k}}$. $\bf{{M_k}}$ is the noise level map. In order to keep the balance between complexity and performance, similar to the parameter settings in \cite{K. Zhang}, the depth of the network is set as 15 and it has a receptive field of $62 \times 62$. The patch size is set to $70 \times 70$ since it should be larger than the receptive field. Then the noisy patches are obtained by adding AWGN to the noise-free patches. The noisy patches and $\bf{{M_k}}$ are input into the CNN. Finally, during the network training, the adaptive moment estimation algorithm is adopted to optimize the FFDNet by minimizing the loss function given as
\begin{align}
L\left( \Theta  \right) = \frac{1}{{2{N_p}}}\sum\limits_{k = 1}^{N_p} {{{\left\| {F\left( {{y_k},{\bf{{M_k}}};\Theta } \right) - {x_k}} \right\|}^2}} .
\end{align}
\section{Simulation Results}
In this section, in order to effectively reflect the accuracy of channel estimation by FFDNet, the simulation results are given to evaluate the peak signal to noise ratio (PSNR) performance of channel estimation achieved with FFDNet in the m-MIMO VLC systems. Moreover, the achieved performance is compared with those obtained with the traditional channel estimation scheme based on MMSE.
\begin{table}[!htbp]
\centering
Table 1: BASIC SIMULATION PARAMETERS
\begin{center}
\begin{tabular}{|c|c|c|}\hline
Description&Notation&Value \\ \hline
The size of the room&${l_r} \times {w_r} \times {h_r}$&$8{\rm{m}} \times 8{\rm{m}} \times 4{\rm{m}}$ \\ \hline
Transmitter semi-angle&${{\varphi _{semi}}}$&${50\,^ \circ }$ \\ \hline
LED power&$P$&$0.02\,{\rm{W}}$ \\ \hline
FOV of the PD&${{\phi _{oc}}}$&${45\,^ \circ }$ \\ \hline
Physical area of PD&${{A_r}}$&$1\,{\rm{c}}{{\rm{m}}^2}$ \\ \hline
Gain of optical filter&${T_s}\left( {{\phi _{ij}}} \right)$&$1$ \\ \hline
Refractive index of OC&$n$&$1.5$ \\ \hline
\end{tabular}
\end{center}
\end{table}

Based on the basic parameters of the m-MIMO VLC system shown in Table 1 \cite{L. Yin}, a number of channel images are obtained by implementing the m-MIMO VLC systems. Some of these images are used as the training data of the FFDNet network and the remaining images are exploited as the testing data. After the training and testing of the network, the performance of FFDNet for the channel estimation of the m-MIMO VLC system is presented and evaluated. Without loss the generality, the size of the transceiver array ${N_t} \times {N_r}$ is set to $128 \times 128$ or $256 \times 256$. And Fig. 4 is given to illustrate the noise level sensitivity curves of FFDNet under both transceiver array sizes.
\begin{figure}[!htbp]
\centering
\subfigure[$128 \times 128$]{
\begin{minipage}{4cm}
\centering
\includegraphics[width=4.3cm]{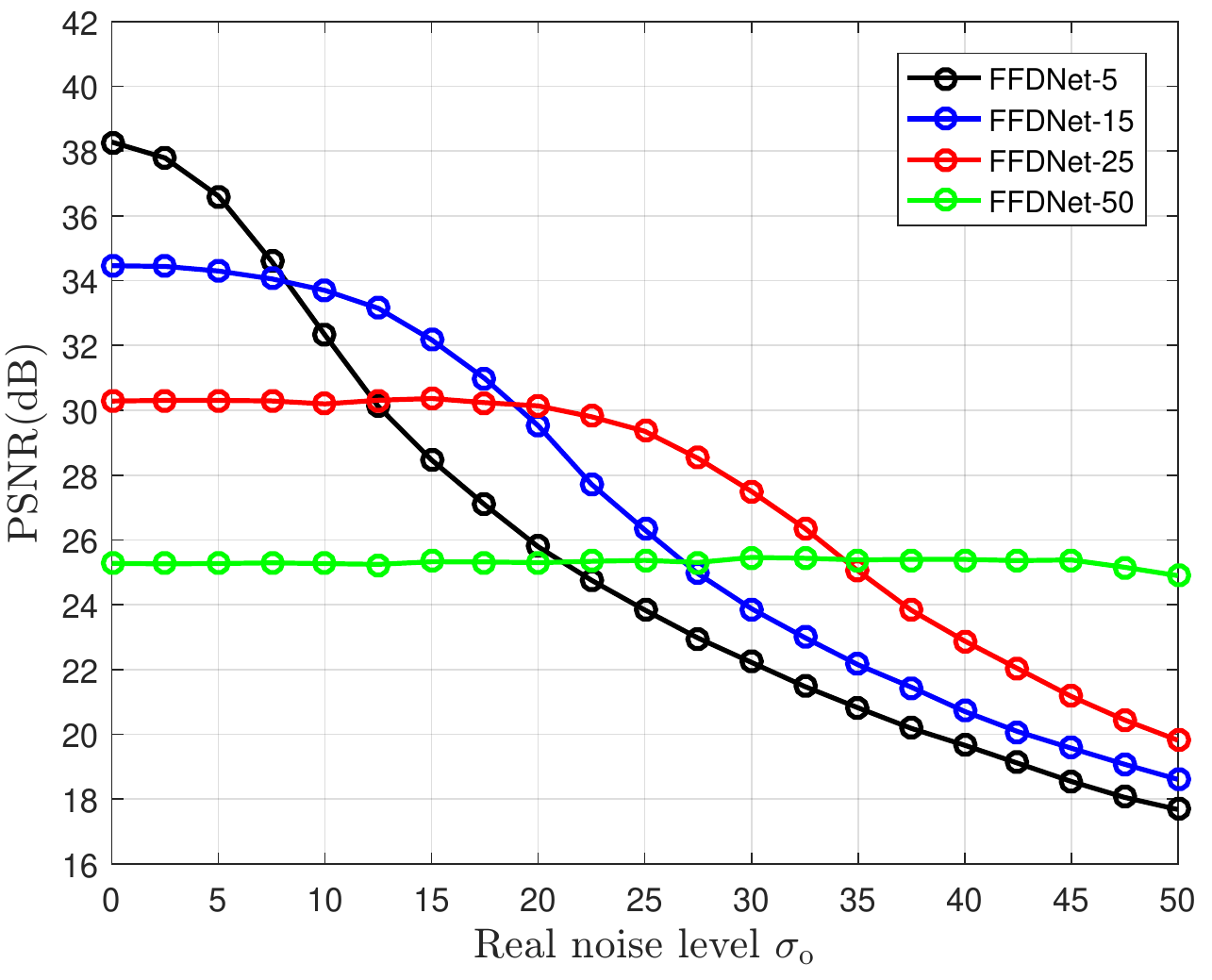}
\end{minipage}
}
\subfigure[$256 \times 256$]{
\begin{minipage}{4cm}
\centering
\includegraphics[width=4.3cm]{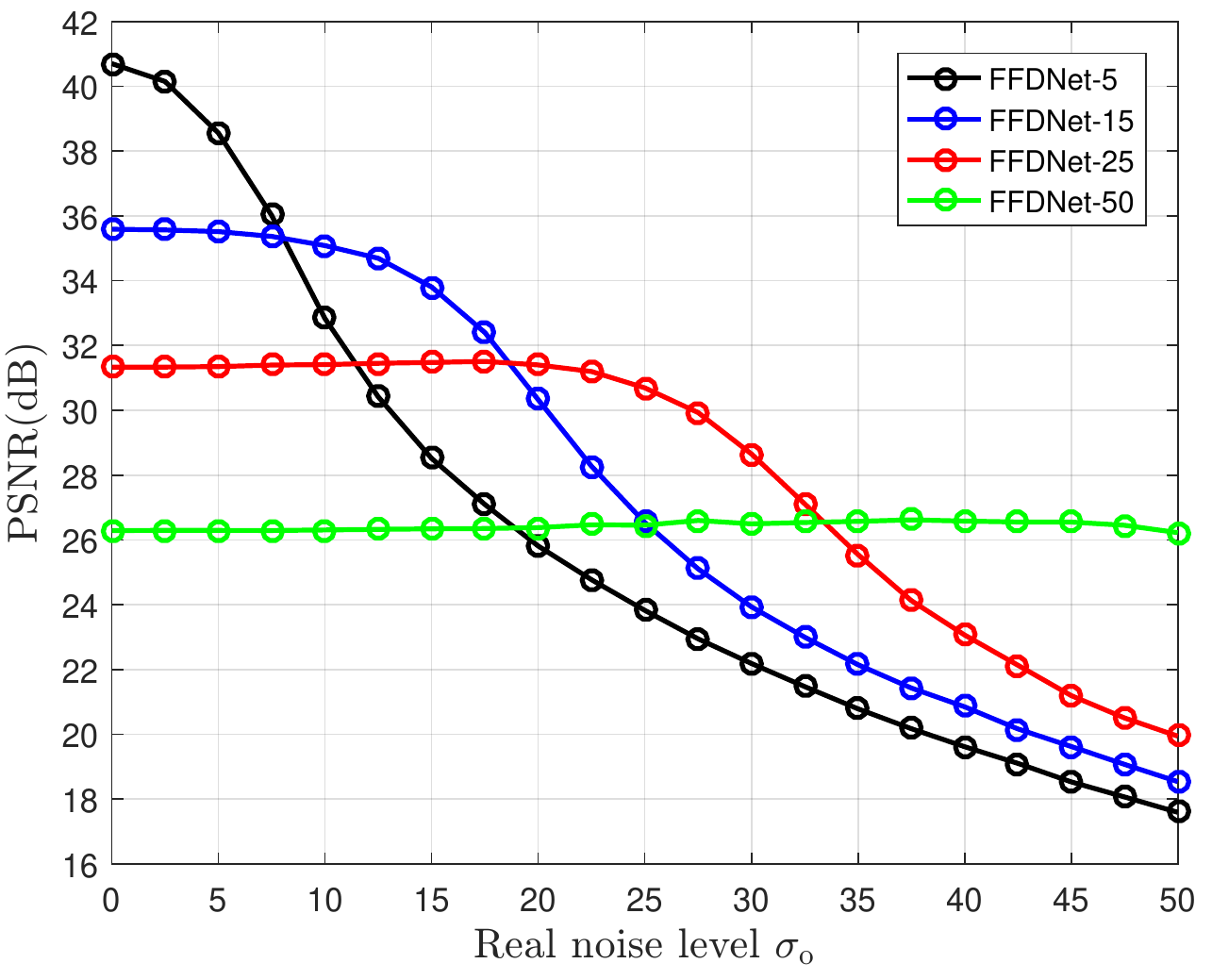}
\end{minipage}
}
\caption{The PSNR performance of channel estimation achieved by FFDNet with different sizes of the transceiver array.} \label{fig.4}
\end{figure}

As shown in Fig. 4, the PSNR is demonstrated by varying different real noise levels ${\sigma _{\rm{o}}} \in \left[ {0,50} \right]$ for the specific noise level maps with $\sigma  = 5,15,25,50$. It can be seen that the PSNR value begins to decrease obviously when the real noise level ${\sigma _{\rm{o}}}$ is larger than the input noise level $\sigma $. And the PSNR is decreased with the increase of ${\sigma _{\rm{o}}}$. Therefore, in practical applications when FFDNet is exploited to estimate the m-MIMO VLC channel, it is better to choose an appropriate $\sigma $ which is larger than the current ${\sigma _{\rm{o}}}$ to achieve a satisfactory noise reduction. Moreover, it can also be seen that the PSNR values for the specific $\sigma $ in Fig. 4 (b) are larger than those for the same specific $\sigma $ in Fig. 4 (a). This is because the number of LEDs and PDs in the transceiver array of the m-MIMO VLC system increases and the array becomes more dense. In this case, the channel matrix has a more obvious sparsity, and its feature information is closer to the 2D natural image. Thus, the channel estimation performance with the FFDNet network based on the image denoising is better.

In order to evaluate the performance of channel estimation with FFDNet in the m-MIMO VLC system, based on the same settings as in Fig. 4 (b), Figs. 5 and 6 are given to compare the performance achieved by FFDNet with those obtained by the traditional channel estimation scheme based on MMSE.
\begin{figure}[!htbp]
\centering
\includegraphics[width=2.0 in]{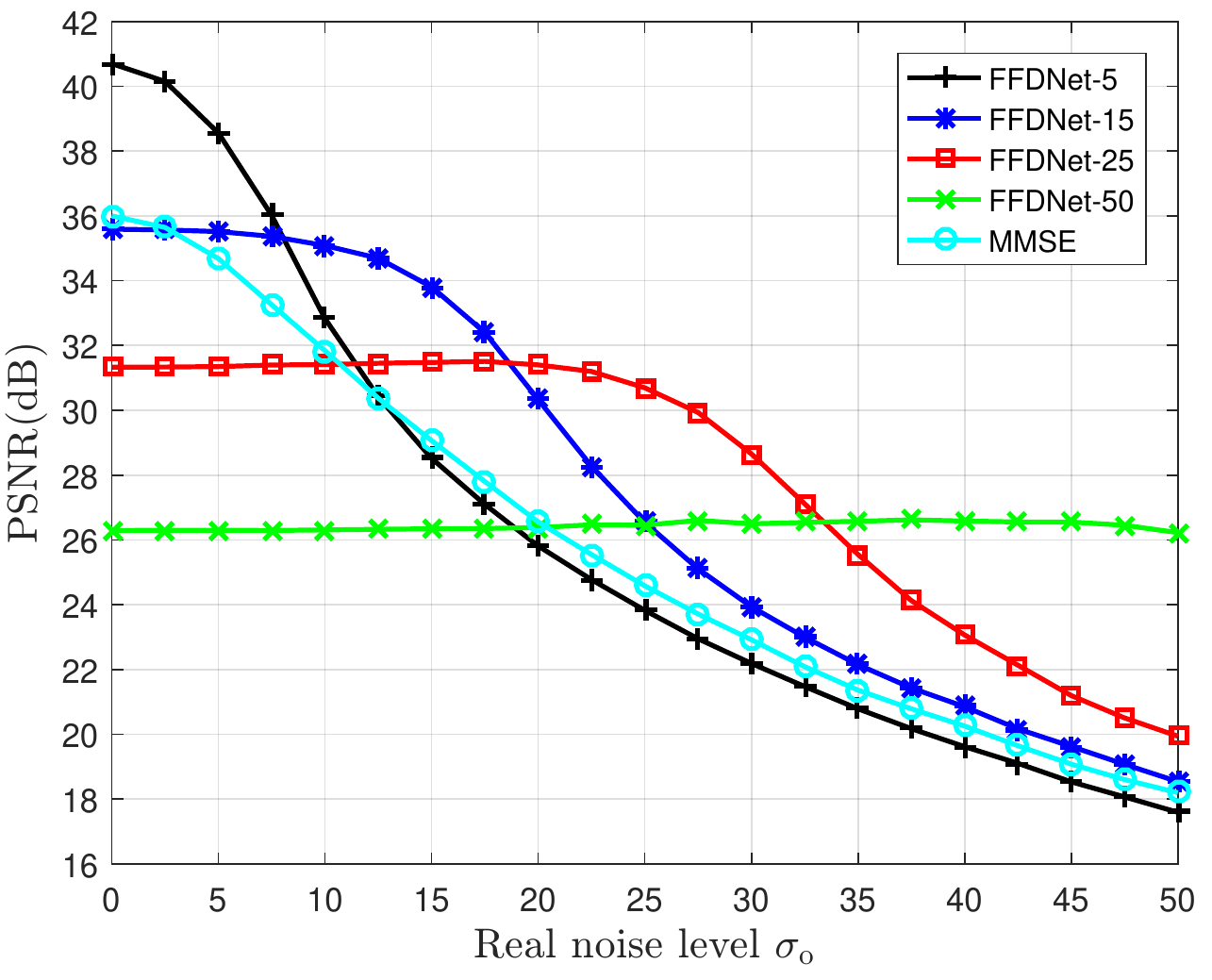}
\caption{The PSNR performance of channel estimation achieved by FFDNet with a fixed $\sigma $ and that obtained by MMSE.} \label{fig.5}

\end{figure}

As shown in Fig. 5, the PSNR performance of the curve named FFDNet-15 is bascically better than that of MMSE under different ${\sigma _{\rm{o}}}$. Although the PSNR values of MMSE are larger than those obtained by FFDNet in the case of the low ${\sigma _{\rm{o}}}$ region, the performance of FFDNet is still better than that of MMSE at high ${\sigma _{\rm{o}}}$. Moreover, the biggest advantage of FFDNet is to exploit a variable $\sigma $ to cope with the interference of different ${\sigma _{\rm{o}}}$. Therefore, we set an appropriate $\sigma $ in FFDNet for the current ${\sigma _{\rm{o}}}$ and compare with the results of MMSE. As shown in Fig. 6, a more satisfactory denoising effect than that of MMSE can be achieved by FFDNet. And the advantages of FFDNet become more obvious as the ${\sigma _{\rm{o}}}$ increases.
\begin{figure}[!htbp]
\centering
\includegraphics[width=2.0 in]{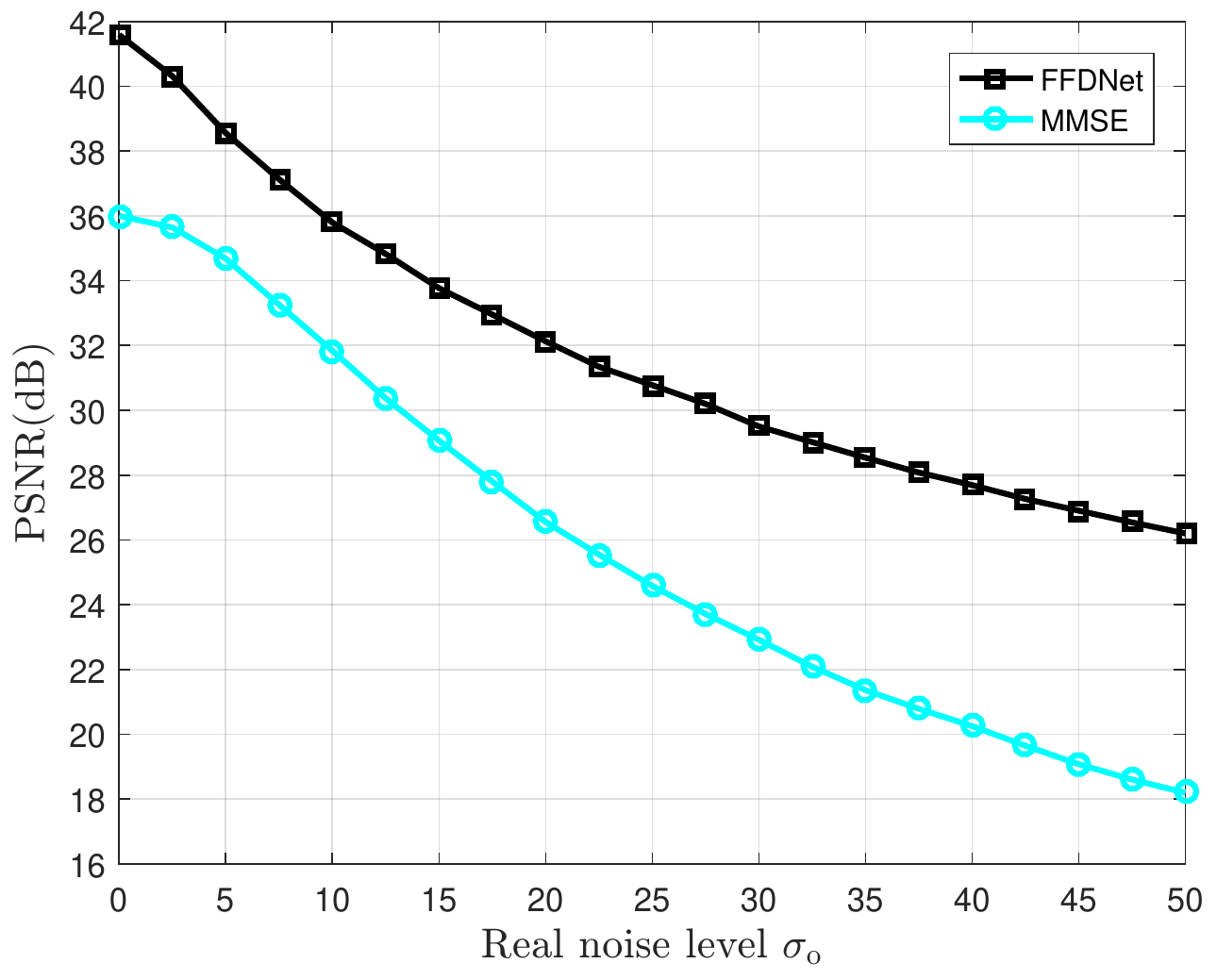}
\caption{The PSNR performance of channel estimation achieved by FFDNet with a tunable $\sigma $ and that obtained by MMSE.} \label{fig.6}
\end{figure}
\section{Conclusion}
A FFDNet-based channel estimation scheme was proposed to estimate the channel of a m-MIMO VLC system. In order to realize it, the channel matrix was established as a 2D natural image. The training method for the FFDNet was provided. Simulation results have shown the superiority of our proposed FFDNet-based channel estimation scheme compared with the existing MMSE scheme.

\end{document}